\documentstyle[prl,aps,epsf,axodraw]{revtex}
\def\br{\begin{eqnarray}}
\def\er{\end{eqnarray}}
\def\be{\begin{equation}}
\def\ee{\end{equation}}

\def\({\left(}
\def\){\right)}

\def\<{\left\langle}
\def\>{\right\rangle}
\def\gc{\<{\frac{\alpha_s}{\pi}}G^{\mu\nu}G_{\mu\nu}\>}
\def\fc{\<{ - \bar{\psi}}\psi\>}
\begin{document}
\twocolumn[\hsize\textwidth\columnwidth\hsize\csname %%% TWO COLUMN
@twocolumnfalse\endcsname                            %%% TWO COLUMN

%
%\draft
%
\title{Relating a gluon mass scale to an infrared fixed point in pure gauge QCD}
\author{A.~C.~Aguilar, A.~A.~Natale and P.~S.~Rodrigues da Silva\\}
\address{Instituto de F\'{\i}sica Te\'orica,
Universidade Estadual Paulista,
Rua Pamplona 145,
01405-900, S\~ao Paulo, SP,
Brazil}
%
%%%
\date{\today}
\maketitle
%%%%

\begin{abstract}
We show that in pure gauge QCD (or any pure non-Abelian gauge
theory) the condition for the existence of a global minimum of energy
with a gluon (gauge boson) mass scale also implies the existence of
a fixed point of the $\beta$ function. We argue that
the frozen value of the coupling constant
found in some solutions of the Schwinger-Dyson equations of QCD can
be related to this fixed point.
We  also discuss how the inclusion of fermions modifies this
property.
\end{abstract}

\pacs{PACS: xxxxxxx}

\vskip 0.5 cm]                            %two column

Non-Abelian gauge theories have the property of asymptotic freedom
\cite{politzer}. For large momenta the coupling becomes small, and
perturbation theory seems to be an appropriate computational tool.
For small momenta the coupling grows large, and we have to rely on
nonperturbative methods to study the infrared (IR) behavior of
these theories. In general, it is easier to apply nonperturbative
methods to pure gauge theories, {\it i.e.} the absence of fermions
may simplify the calculations. One of these methods, in the case
of pure gauge quantum chromodynamics, is the study of
Schwinger-Dyson equations (SDE) for the gluon propagator
\cite{mandelstam}. Following this method it was found some years
ago that the gluon propagator is highly singular in the IR, which
could explain gluon confinement in a simple way \cite{mandelstam}.
This early calculation contained a series of approximations, and
nowadays it is believed that the gluon propagator IR behavior is
smoother.

The softer IR behavior of the gluon propagator indicates the
existence of a gluon mass scale. This conclusion was reached by a
large number of nonperturbative methods. Cornwall argued that the
gluon acquires a dynamical mass solving a gauge invariant SDE
\cite{cornwall}. Recent research using a similar method with
different approximations also finds an IR finite propagator
involving a gluon mass scale \cite{alkofer}. These calculations
are consistent with lattice simulations of pure gauge QCD, where
it is found that the gluon propagator is modified at some mass
scale and is infrared finite \cite{lat}. A variational method
approach to QCD is also compatible with dynamical gluon mass
generation \cite{kogan}. This gluon mass scale appears in some
other nonperturbative methods \cite{various}, as well as is
necessary in several phenomenological calculations \cite{halzen}.

At the same time that the dynamical gluon mass scale is generated,
the theory develops a freezing of the IR coupling constant. This
is a consequence that the coupling behavior is related to the
renormalization of the theory propagators, and in this procedure
the infrared behavior of the gluon is transmitted to the coupling.
Actually the coupling constant found in Ref.\cite{alkofer},
solving SDE, clearly shows the existence of an infrared fixed
point of the QCD $\beta$ function. Nevertheless, it is important
to stress that the SDE solutions are always solved within some
approximation and in general in one specific gauge and
renormalization scheme. Therefore we expect that any relationship
between the gluon mass scale and the infrared behavior of the
coupling constant and, consequently, a fixed point of the $\beta$
function could not be univocally determined. This fact is peculiar
to our inability to deal with the strong interaction physics,
because we expect that the absolute minimum of QCD vacuum energy
will be compatible with a unique gluon mass scale (if this is the
solution preferred by the vacuum).

In this work we show that the dynamical gluon mass scale
generation implies the existence of a fixed point of the $\beta$
function, although the presence of a fixed point does not
necessarily imply dynamical mass generation. We start remembering
that gauge theories without fundamental scalar bosons may generate
dynamical masses through the phenomenon of dimensional
transmutation~\cite{coleman}; {\it i.e.}, we basically do not have
arbitrary parameters once the gauge coupling constant $(g)$ is
specified at some renormalization point $(\mu)$. In these theories
all the physical parameters will depend on this particular
coupling.

Many years ago Cornwall and Norton~\cite{cornwall2} emphasized that the
vacuum energy $(\Omega)$ in dynamically broken gauge
theories could be defined as a function of the dynamical mass
$m_g (p^2) \equiv m (g,\mu)$,
%
%%%\begin{equation}
%%%m_g (p^2) \equiv m (g,\mu) \; ,
%%%\label{e1}
%%%%%\end{equation}
%
where $m_g (p^2)$ in pure gauge QCD is related to the gluon
polarization tensor. This mass is not necessarily the gluon mass
as it appears in the Euclidean propagator determined in
Ref.\cite{cornwall}, it may be any momentum dependent mass scale
that induces an IR finite behavior for the gluon propagator as it
appears in Ref.\cite{alkofer}. In the sequence $m (g,\mu)$ will be
indicated just by $m$.  Actually the vacuum energy may also depend
on the dynamical fermion and ghost masses. However, in that which
concerns ghosts,  there is no evidence for scalar fermion
Goldstone excitations, {\it i.e.} it is rather unlikely that
ghosts develop mass \cite{eichten}.

The vacuum energy $\Omega = \Omega (g,\mu)$, defined ahead, is a
finite function of its arguments, because the perturbative
contribution has been subtracted \cite{cornwall2,cjt}. $\Omega$
must satisfy a homogeneous renormalization group
equation~\cite{gross}
\begin{equation}
\left( \mu \frac{\partial}{\partial \mu} + \beta (g) \frac{\partial}{\partial g}
\right) \Omega = 0 \; .
\label{e2}
\end{equation}
On the other hand, the dynamically generated masses
can be written as $m =\mu f(g)$~\cite{gross},
from what follows that $\mu (\partial m / \partial \mu ) = m$
and, consequently,
\begin{equation}
m \frac{\partial \Omega}{\partial m} = - \beta(g)
\frac{\partial \Omega}{\partial g} \; .
\label{e3}
\end{equation}
This last and simple equation will be central to our argument,
because it relates the stationary condition for the vacuum energy
$(\partial \Omega / \partial m = 0)$~\cite{cornwall,cjt} to the
condition of zeros of the $\beta$ function, and we expect that the
massive solution indeed minimizes the energy \cite{kogan}.
Therefore in a gauge theory with dynamically generated masses, the
condition for an extremum of the vacuum energy:
\begin{equation}
\beta(g) \left. \frac{\partial \Omega}{\partial g}
\right|_{\partial \Omega / \partial m = 0} = 0 \;
\label{e4}
\end{equation}
always implies $\beta(g)=0$. Of course, this is true only if
$\partial \Omega / \partial g \neq 0$ when $m \neq 0$. Note that
only at the global minimum is the vacuum energy  a gauge
independent and meaningful quantity. Exactly at this point we
expect that the mass scale, the coupling constant, and its $\beta$
function are uniquely determined.

We note that the coupling constant in the IR has no unique
determination, and it has been enough to match its functional form
with its ultraviolet behavior. This diversity at the IR has the
inconvenience that depending on the choice we make, we have to
face very different scenarios, for instance, the singular behavior
of the coupling\cite{bode} or its freezing at low
energies\cite{cornwall,alkofer}. In this sense, it would be
appropriate to clarify what coupling constant and $\beta$ function
we are referring to, since Eq.(\ref{e3}) was written down without
any specification of their functional form and the renormalization
scale where they are to be computed. The point here is that
Eq.(\ref{e3}) precedes any  {\it a priori} definition of the
coupling constant and its associated $\beta$ function (at some
renormalization scale), allowing us to obtain very general
properties of these functions if we have some extra ingredient at
hand.  As discussed by Coleman and Weinberg many years ago
\cite{coleman}, there is a unique way of linking $g$ and $\mu$ and
this can be achieved by the vacuum energy at its minimum. In other
words, whatever the definition of $g$ and $\beta$ we choose, the
minimum of energy provides us with further information, demanding
them to conform to the existence of an IR fixed point when there
is dynamical mass generation.

To show that $\partial \Omega / \partial g \neq 0$ we must
refer to the vacuum energy for composite operators \cite{cjt},
since the theory will admit only condensation of composite
operators as, for instance, $\gc$ in the pure gauge theory and
$\fc$ when we add fermions. In order to do so we introduce a
bilocal field source $J(x,y)$, and $\Omega$ will be calculated
after a series of steps starting from the generating functional
$Z(J)$~\cite{cjt}:
\begin{eqnarray}
Z(J) &=& \exp [\imath W(J)]= \int d\phi
\exp \left[ \imath \left( \int d^4 x {\cal L}(x) \right. \right.\nonumber \\
&& + \left. \left. \int d^4x d^4y {\phi}(x)J(x,y){\phi}(y) \right) \right] \; ,
\label{e5}
\end{eqnarray}
where $\phi$ can be a gauge boson or fermion field.
From the generating functional we determine the effective action $\Gamma(G)$
which is a Legendre transform of $W(J)$ and is given by
$\Gamma(G)=W(J)- \int d^4x d^4y G(x,y)J(x,y)$
%
%%%\begin{equation}
%%%\Gamma(G)=W(J)- \int d^4x d^4y G(x,y)J(x,y) \; ,
%%%\label{e6}
%%%\end{equation}
%
(where $G$ is a complete propagator) leading to
$\delta \Gamma / \delta G(x,y)=-J(x,y)$.
%
%%%\begin{equation}
%%%\delta \Gamma / \delta G(x,y)=-J(x,y) \; .
%%%\label{e7}
%%%\end{equation}
%
The physical solutions will correspond to $J(x,y)=0$, which will
reproduce the SDE of the theory~\cite{cjt}.

In general, if $J$ is the source of the operator $\cal{O}$, we have~\cite{gkm}
\begin{equation}
\left. \frac{\delta \Gamma}{\delta J } \right|_{J=0}=
\left\langle 0 \left| \cal{O} \right| 0 \right\rangle \; .
\label{e8}
\end{equation}
For translationally invariant (ti) field configurations we can
work with the effective potential given by  $V(G) \int d^4x = -
\Gamma(G)|_{ti}$.
%
%%%%\begin{equation}
%%%%V(G) \int d^4x = - \Gamma(G)|_{t.i.} \; .
%%%%%\label{e9}
%%%%%%\end{equation}
%
Finally, from the above equations we can define the vacuum energy as~\cite{cjt}
\begin{equation}
\Omega = V(G) - V_{pert}(G) \; ,
\label{e10}
\end{equation}
where we are subtracting from $V(G)$ its perturbative counterpart,
and $\Omega$ is computed as a function of the nonperturbative
propagators $G$.  These propagators depend on the gauge boson,
fermion and ghosts self-energies. We will not consider fermions
and, as long as the ghost self-energy does not show any nontrivial
pole, its direct contribution is washed out from the vacuum
energy. It should be noted, however,  that the ghosts can still
interfere through its effect on the gluon propagator
\cite{alkofer}. $\Omega$ is a function of the dynamical masses of
the theory and is zero in the absence of mass generation
\cite{cjt}. We shall comment later on the actual $\Omega$
calculation.

We can now write Eq.(\ref{e4}) in the following form:
\begin{equation}
 - \beta(g) \left[ \frac{\partial \Omega}{\partial J}
\frac{\partial J}{\partial g} \right]_{J=0} = 0 \; .
\label{e13}
\end{equation}
Of course, we assume that the conditions for a global minimum of
the vacuum energy $\partial \Omega / \partial m = 0$ and $J=0$ are
equivalent. However, $\partial \Omega / \partial J = - \partial
\Gamma /
\partial J$, and as a consequence of Eq.(\ref{e8}) we have
\begin{equation}
\beta(g) \left\langle 0 \left| \cal{O} \right| 0 \right\rangle
\left. \frac{\partial J}{\partial g} \right|_{J=0} = 0 \; .
\label{e14}
\end{equation}

Using the inversion method devised by Fukuda \cite{fukuda} it is
possible to show that $\frac{\partial J}{\partial g} \neq 0$ when
there is condensation, i.e., $\left\langle 0 \left| \cal{O}
\right| 0 \right\rangle \equiv \vartheta\neq 0$. In
Ref.\cite{fukuda} it was verified that to compute a
nonperturbative quantity like $\vartheta$ the usual procedure is
to introduce a source $J$ and to calculate the series
\begin{equation}
\vartheta = \sum_{n=0}^{\infty}  g^n h_n(J) \; .
\label{e15}
\end{equation}
In practice we have to truncate Eq.(\ref{e15}) at some finite
order, which gives us only the perturbative solution $\vartheta
=0$ when we set $J=0$. The right-hand side of Eq.(\ref{e15})
should be double valued at $J=0$ for another solution to exist,
which is not the present case. The alternative method is to invert
Eq.(\ref{e15}), solving it in favor of $J$ and regarding
$\vartheta$ as a quantity of the order of unity. One obtains the
following series:
\begin{equation}
J=  \sum_{n=0}^{\infty} g^n k_n (\vartheta) \; ,
\label{e16}
\end{equation}
where the $k_n$'s satisfying $n \leq m$ ($m$ being some finite
integer) are calculable from $h_n$ also satisfying $n \leq m$. One
can find a nonperturbative solution of $\vartheta$ by setting
$J=0$ through a truncated version of Eq.(\ref{e16}). The important
point for us is that by construction of Eq.(\ref{e16}) we verify
that when $J=0$ and $\vartheta \neq 0$ the same value of
$\vartheta$ that satisfies Eq.(\ref{e16}) leads trivially to
\begin{equation}
\partial J / \partial g |_{J=0} \neq 0 \; .
\label{e16a}
\end{equation}

To make this point clear, observe that Eq.(\ref{e16}) allows us to
look at $J$ as a function of $g$ and $\vartheta$; hence we can
imagine a surface in the space spanned by $J$, $g$ and
$\vartheta$. Nevertheless, this surface has physical meaning only
for $J=0$, resulting in a curve in the ($g$,$\vartheta$) plane
where the derivative of Eq.(\ref{e16a}) is calculated. Therefore,
the two terms, $\partial J /
\partial g$ and $\left\langle 0 \left| \cal{O} \right| 0
\right\rangle$, of Eq.(\ref{e14}) are different from zero in
the condensed phase.

According to the above discussion and looking at Eq.(\ref{e14}),
the only possibility to obtain $\partial \Omega / \partial m = 0$
is when we have a fixed point [$\beta (g) = 0$], from which comes
our main assertion that the condition for the existence of a gluon
(gauge boson) mass scale at the global minimum of the vacuum
energy also implies the existence of a fixed point of the $\beta$
function. The reverse is not necessarily true, since the theory
may have a fixed point consistent with the absence of any
dynamical mass.

It should be remembered that there are more than one SDE solution
consistent with a dynamical mass scale for the gluon. These
solutions, as discussed previously, depend on the different
approximations used to solve the equations and gauge choice, and
they necessarily do not lead to a global minimum of energy. It is
reasonable to expect that only  the true solution, massive or not,
will give the absolute minimum of energy and if it has a gluon
mass scale it will be related to a unique fixed point.

We can demonstrate the connection between the gauge boson mass
scale and the existence of the fixed point in a different way if
we particularize the problem to pure gauge QCD. Its Lagrangian is
given by ${\cal L}= \frac{1}{2}G_{\mu\nu}^2$ and $\phi= A_\mu$ in
Eq.(\ref{e5}). Following an argumentation presented by
Cornwall~\cite{cornwall} we can now rescale the fields $A_\mu^a
=g^{-1}{\hat{A}}_\mu^a$, $G_{\mu\nu}^a =g^{-1}
{\hat{G}}_{\mu\nu}^a$, and regularize the vacuum energy (and the
potential) setting its perturbative part equal to zero in order
to obtain
\begin{eqnarray}
Z &=& Z_p^{-1} \int d{\hat{A}}_\mu
\exp \left[ -g^{-2} \int d^4 x\frac{1}{4} \< \sum_a ({\hat{G}}_{\mu\nu}^a)^2
\>\right]\nonumber \\
&& = e^{-V \Omega} \; ,
\label{e11}
\end{eqnarray}
where $V$ is the volume of Euclidean space-time and $Z_p$ is the perturbative functional.
Differentiating with respect to $g$ it follows that
\begin{equation}
\frac{\partial \ln Z}{\partial g} = \frac{1}{2g} \int d^4 x
\< \sum_a ({\hat{G}}_{\mu\nu}^a)^2 \>_{reg} = - \frac{V\partial\Omega}{\partial g},
\label{e12}
\end{equation}
where the subscript ''reg" on the gluon condensate indicates that
the regularization is by subtraction of the perturbative
expectation value in the same way as indicated in Eq.(\ref{e10}).
The factor $V$ in the right-hand side is canceled with the one
coming out from the $x$ integration. As long as the condensate is
different from zero for some $g>0$, and there are indications that
this happens for any $g>0$~\cite{fukuda1} (and the same would
happen for the gluon mass scale\cite{cornwall,alkofer}), $\partial
\Omega /
\partial g \neq 0$, since this quantity is proportional to the
condensate. This argumentation is correct only in the light-cone
gauge (or any ghost-free gauge) as discussed in
Ref.\cite{cornwall}, for which the derivation of Eq.(\ref{e12}) is
valid. Furthermore, the condensate must be consistent with the
deepest minimum of energy. According to Eq.(\ref{e3}) and
Eq.(\ref{e4}), this result constitutes an alternative proof of our
statement that the theory has a nontrivial fixed point at the
global minimum of energy, though restricted to a particular
scheme.

It would be suitable to compute the vacuum energy $\Omega$ and
show explicitly the connection between its minimum and the fixed
point. However, to compute $\Omega$  we must know the full
nonperturbative Green functions of the theory, which obviously is
not an easy task. In general this is accomplished using IR
finite propagators within some rough approximations
\cite{cornwall,kogan,montero}.

We can now discuss what happens if instead of a pure gauge theory
we also have fermions. Actually part of the arguments presented
here were already discussed by some of us when studying fermionic
condensation and mass generation in the case of strong coupling
QED \cite{natale}, but the implications were not fully realized
and only later it became clear to us \cite{gorbar} that the vacuum
energy in QCD with massless fermions is basically dominated by the
gluonic (gauge boson) condensation (or mass) rather than by the
fermionic one. This fact can be observed if we recover some of the
results of Ref.\cite{gorbar} in the following form:
\begin{equation}
\< \Omega \> \propto - \frac{1}{16\pi^2} \left[ \frac{3(N^2
-1)}{2}a m^4 +N b\eta^4 \right] , \label{e18}
\end{equation}
where $\< \Omega \>$ is the QCD vacuum energy at the extrema of
energy in the case that we have a massless fermion, $N=3$ is the
number of colors, $a$ and $b$ are constants determined by the
theory and calculated in Ref.\cite{gorbar}, $m$ is the gluon mass
and $\eta$ is the dynamical fermion mass, which are ultimately
connected to the gluon and fermion condensates. There are several
points to discuss about this expression. First, it was derived in
Landau gauge and involves many approximations. We are far from a
satisfactory determination of the full momentum dependence of the
dynamical masses used as input to compute Eq.(\ref{e18}), but we
believe that this equation can roughly describe the actual
behavior. Second, currently assumed values for the gluon and
fermion masses \cite{halzen},\cite{montero} -\cite{gorbar}
indicate that the first term of the right hand side dominates the
other by at least 1 order of magnitude. Usual estimates of the
dynamical masses give the ratio $m/\eta \approx 2$. Third, as is
well known, the gluonic SDE are coupled to the fermionic and the
ghost ones, i.e., the dynamical gauge boson mass is affected by
the presence of fermions and ghosts and vice versa. However, in
the case of fermions the effect is small, at least in what
concerns the gluon mass \cite{gorbar,aguilar}. Therefore, the
global minimum of energy of QCD (or any other non-Abelian gauge
theory) is dictated by the gauge bosons, and we can argue that any
fixed point of the theory will be determined by the gauge boson
sector. The fermions introduce only small changes in the position
of the vacuum energy. It is also clear that if we increase the
number of fermions too much we will change the values of the
dynamical masses as well as the relative importance of each term
in Eq.(\ref{e18}).

In conclusion, we have shown that the condition for the global
minimum of the vacuum energy for a non-Abelian gauge theory with a
dynamically generated gauge boson mass scale implies the existence
of a nontrivial IR fixed point of the theory. This vacuum energy
depends on the dynamical masses through the nonperturbative
propagators of the theory. Our results show that the freezing of
the QCD coupling constant observed in the calculations of
Ref.~\cite{cornwall,alkofer} can be a natural consequence of the
onset of a gluon mass scale, giving strong support to their claim.
\section*{Acknowledgments}
This research was partially supported by the Conselho Nacional de
Desenvolvimento Cient\'{\i}fico e Tecnol\'ogico (CNPq) (A.A.N.)
and by Funda\c{c}\~ao de Amparo \`a Pesquisa do Estado de S\~ao
Paulo (FAPESP) (A.C.A. and P.S.R.S.).

\end{document}